\documentclass[pre,aps,preprint,amsmath]{revtex4}

\usepackage{fancyhdr}
\usepackage{graphicx}
\usepackage{dcolumn}
\usepackage{txfonts}
\usepackage{bm}
\usepackage{mathrsfs}
\usepackage{amsfonts}
\usepackage{amssymb}
\usepackage{braket}
\usepackage{color}
\usepackage[colorlinks=true,citecolor=magenta,anchorcolor=blue]{hyperref}

\begin{document}

\title{Topological thermal transport}

\author{Zhoufei Liu}
\author{Peng Jin}
\author{Min Lei}
\author{Chengmeng Wang}
\affiliation{Department of Physics and State Key Laboratory of Surface Physics, Fudan University, Shanghai 200438, China}

\author{Fabio Marchesoni}
\affiliation{MOE Key Laboratory of Advanced Micro-Structured Materials and Shanghai Key Laboratory of Special Artificial Microstructure Materials and Technology, School of Physics Science and Engineering, Tongji University, Shanghai 200092, China\\
Department of Physics, University of Camerino, 62032 Camerino, Italy}

\author{Jian-Hua Jiang}\email{jhjiang3@ustc.edu.cn}
\affiliation{Suzhou Institute for Advanced Research, University of Science and Technology of China, Suzhou, Jiangsu, 215123, China} 

\author{Jiping Huang}\email{jphuang@fudan.edu.cn}
\affiliation{Department of Physics, State Key Laboratory of Surface Physics, and Key Laboratory of Micro and Nano Photonic Structures (MOE), Fudan University, Shanghai 200438, China}

\date{\today}

\begin{abstract}

Thermal transport is a fundamental mechanism of energy transfer process quite distinct from wave propagation phenomena. It can be manipulated well beyond the possibilities offered by natural materials with a new generation of artificial metamaterials: thermal metamaterials. Topological physics, a focal point in contemporary condensed matter physics, is closely intertwined with thermal metamaterials in recent years. Inspired by topological photonics and topological acoustics in wave metamaterials, a new research field emerged recently, which we dub `topological thermotics', which encompasses three primary branches: topological thermal conduction, convection, and radiation. For topological thermal conduction, we discuss recent advances in both 1D and higher-dimensional thermal topological phases. For topological thermal convection, we discuss the implementation of thermal exceptional points with their unique properties and non-Hermitian thermal topological states. Finally, we review the most recent demonstration of topological effects in the near-field and far-field radiation. Anticipating future developments, we conclude by discussing potential directions of topological thermotics, including the expansion into other diffusion processes such as particle dynamics and plasma physics, and the integration with machine learning techniques.

\end{abstract}

\maketitle

\section*{Introduction} 

Numerous topological states of matter have been reported over the years\textsuperscript{\cite{HasanRMP10, QiRMP11}}, including quantum spin Hall insulators\textsuperscript{\cite{KanePRL05, BernevigSci06, KonigSci07}}, quantum anomalous Hall insulators\textsuperscript{\cite{YuSci10, ChangSci13}}, topological superconductors\textsuperscript{\cite{KitaevAP03, SatoRPP17}}, Dirac\textsuperscript{\cite{WangPRB12, LiuSci14}} and Weyl semimetals\textsuperscript{\cite{WanPRB11, BurkovPRL11, LvPRX15, ArmitageRMP18}}, to name but a few. Topological materials have been nearly completely catalogued based on the band symmetry analysis\textsuperscript{\cite{VergnioryNat19, TangNat19, ZhangNat19}}. 

An alternative platform for investigating topological physics has arisen in the form of wave metamaterials\textsuperscript{\cite{ZhangNat23, NiCR23}}. Metamaterials are artificial materials with special properties beyond natural materials; wave metamaterials manipulate wave propagation. The interplay between topology and wave dynamics has spawned new areas, two examples being topological photonics\textsuperscript{\cite{MiriSci19, LuNP14, OzawaRMP19}} and topological acoustics\textsuperscript{\cite{XueNRM22, MaNRP19, ZhuRPP23}}. Topological photonics has revealed topological phases elusive in condensed matter physics, such as Floquet topological insulators\textsuperscript{\cite{RechtsmanNat13, MaczewskyNC17}}, non-Hermitian topological states\textsuperscript{\cite{ZhaoSci19, WeidemannSci20}} and topological bulk-defect correspondence\textsuperscript{\cite{LiNC18, LiuNat21-1, LinNRP23}}. Furthermore, topological photonics has opened up new technological applications such as topological lasers\textsuperscript{\cite{HarariSci18, BandresSci18}}. At the same time, topological acoustics has become prominent in the study of topological physics with the discovery of intriguing topological effects\textsuperscript{\cite{YangPRL15}}, such as topological negative refraction\textsuperscript{\cite{HeNat18}}, topological `sasers' (that is, the phononic analog of lasers)\textsuperscript{\cite{HuNat21}}, higher-order topological insulating\textsuperscript{\cite{ZhangNP19, ZhangNC19, XieNRP21}} and semimetal states\textsuperscript{\cite{LuoNM21, XiangPRL24}}, novel topological spectral flows\textsuperscript{\cite{LinNM22}}, and non-Abelian topological phases\textsuperscript{\cite{JiangNP21, QiuNC23}}. These two research fields, together with other fields such as topological mechanics\textsuperscript{\cite{MaNRP19, ZhuRPP23, HuberNP16}}, have enriched the spectrum of topological physics. 

Another mechanism of energy transport is thermal diffusion. Thermal metamaterials have been suggested to provide unprecedented control of heat transport\textsuperscript{\cite{ZhangNRP23, YangRMP24, Yang24, Huang20, Xu23, YangPR21, LiNRM21, JuAM22}}. Thermal cloaking was the first breakthrough effect realized via thermal metamaterials\textsuperscript{\cite{FanAPL08, ChenAPL08, NarayanaPRL12, Yeung22, LouCPL23}}, then followed by recent developments like thermal illusions\textsuperscript{\cite{HuAM18, YangESEE19, HeCPL23}} and thermal coding\textsuperscript{\cite{GuoAM22, YangPRAP23}}. The concepts of thermal metamaterials have been generalized from heat conduction to heat convection; doing so yields new effects and applications\textsuperscript{\cite{DaiJAP18, JinPNAS23}}. By the same token, the third mechanism of heat transfer, thermal radiation, has also found its place in thermal metamaterial technology\textsuperscript{\cite{RamanNat14, ZhaiSci17, XuESEE20, XuPRAP20, ZhangCPL23, YinCPL23}}. 

Thermal metamaterials have been demonstrated to provide a flexible platform to realize various topological phases and topological effects. The topic is gaining traction in the thermal metamaterial literature. In view of the role and development of topological photonics and topological acoustics in wave metamaterials\textsuperscript{\cite{ZhangNat23, NiCR23, MiriSci19, LuNP14, OzawaRMP19, XueNRM22, MaNRP19, ZhuRPP23}}, we expect that their thermal counterpart, which we refer to as ``topological thermotics'', will grow soon into a new interdisciplinary research area. This is the topic of this Perspective. 

According to the three basic heat transfer modes, topological thermotics can be divided up into three primary branches: topological thermal conduction, topological thermal convection and topological thermal radiation. The development of dynamic, convective and hybrid thermal metamaterials supports thermotics as an alternative adjustable platform for topological physics. Owing to the intrinsically dissipative character of diffusion, topological thermotics is already revealing physical phenomena without wave counterparts. Topological thermotics thus opens new venues in nonequilibrium thermodynamics and stimulates further advances in thermal metamaterials research. Indeed, besides its expected impact in fundamental research, topological thermotics has potential also in technological applications that require efficient heat dissipation and localized heat management, from thermal engineering to quantum technologies.

\begin{figure}[!ht]
\includegraphics[width=\linewidth]{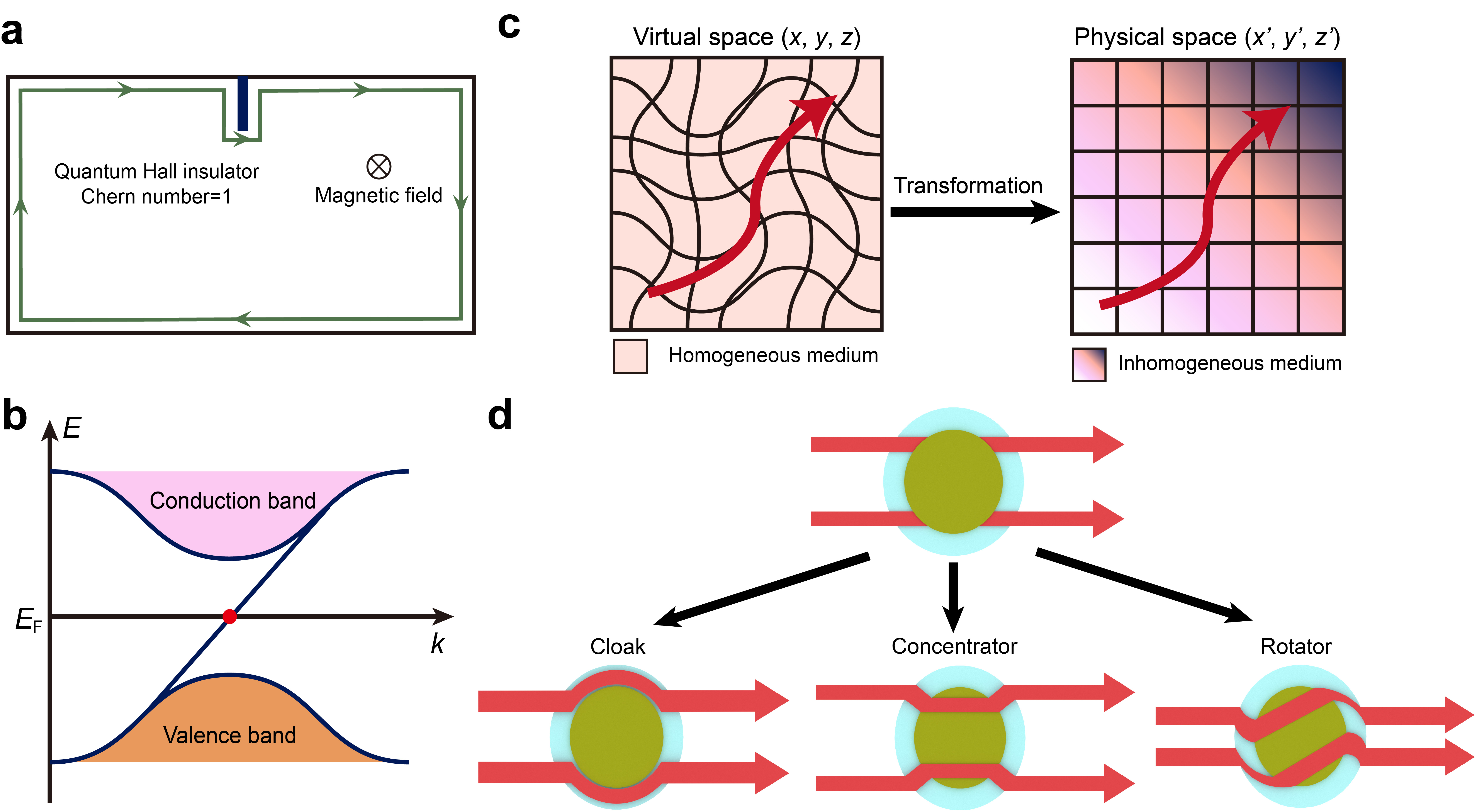}
\caption{{\bf Topological theory and transformation theory.} {\bf a--b}, Topological theory. Schematic of the quantum Hall insulator (part {\bf{a}}). The blue arrow denotes the robust chiral edge state. Band structure, $E(k)$, of the quantum Hall insulator with $E_{\rm{F}}$ denoting the Fermi energy and $k$ the Bloch vector (part {\bf{b}}). {\bf c--d}, Transformation theory. Schematic of the transformation theory (part {\bf{c}}). Wavy arrows represent the relevant heat flow and the grids the coordinate systems. The arrow is curved in the twisted virtual space with homogeneous medium (left). After transforming the virtual space to a Cartesian real space with an inhomogeneous medium, the flow arrow is still curved (right). Applications of transformation theory: cloak, concentrator, and rotator (part {\bf{d}}). The shell areas are the transformed region.}
\label{Fig1}
\end{figure}

Topological thermotics studies the topological physics in thermal metamaterials. So the related theories used in topological thermotics research include topological theory and thermal metamaterial theory. Topological theory is concerned with physical properties that are invariant under continuous deformation\textsuperscript{\cite{Nakahara03}}. In a commonly used example, a doughnut can be continuously deformed into a coffee cup, because both have the same number of holes. However, a spoon cannot be continuously deformed into the doughnut. In this case, the topological invariant corresponds to the number of holes in the closed surface, which is also called genus. The most typical example of topology in condensed matter physics may be the integer quantum Hall effect in a 2D electron gas (Fig.~\ref{Fig1}a). The corresponding topological invariant is Chern number calculated by the integration of Berry curvature over the Brillouin zone\textsuperscript{\cite{ThoulessPRL82, Asboth16}}. The nonzero Chern number predicts the existence of chiral edge states due to the bulk--boundary correspondence (Fig.~\ref{Fig1}a). These edge states traverse the whole band gap (Fig.~\ref{Fig1}b) and cannot be eliminated without closing the band gap. The fundamental theories of thermal metamaterials are the transformation theory and its extended theories\textsuperscript{\cite{YangRMP24}}. Transformation theory, originating from transformation optics\textsuperscript{\cite{PendrySci06, LeonhardtSci06}}, assumes the form-invariance of the governing equations under coordinate transformation. This form invariance can be used to control heat flows by transforming the material parameters (Fig.~\ref{Fig1}c). By such means, many desired thermal functional devices can be designed, such as cloak, concentrator, and rotator (Fig.~\ref{Fig1}d). Many extended theories have been formulated to assist the transformation theory, such as scattering cancellation theory\textsuperscript{\cite{HanPRL14, XuPRL14}}, effective medium theory\textsuperscript{\cite{DaiPRE18}}, pseudo-conformal mapping theory\textsuperscript{\cite{DaiCSF23}} and conformal diffusion tracing theory\textsuperscript{\cite{XuNCS24}}. In particular, the interplay between topological theory and transformation theory with its extended theories enriches the scope of topological thermotics. For example, a flexible switch between cloaking and concentration can be achieved by applying the transformation theory to the conduction--convection equation, the possibility of which points to a new topological transition\textsuperscript{\cite{JinPNAS23}}.

The experimental methods of topological thermotics are inherited from those for thermal metamaterials. In the fabrication of materials, the common strategies include machining\textsuperscript{\cite{HuAM22, WuAM23}} and 3D printing\textsuperscript{\cite{QiAM22}}. Effective medium theory opens a practical pathway towards inhomogeneity\textsuperscript{\cite{NarayanaPRL12}} and anisotropy\textsuperscript{\cite{TianIJHMT21}} of material parameters. Thermal convection can be included by introducing liquid elements in the metamaterials. Driving the motion of the solid and/or liquid elements gives rise to dynamically modulated heat flows which can be controlled by motors and steering gears\textsuperscript{\cite{XuNP22, XuNC23}}. All these experimental advances have enriched the study of thermal metamaterials, and may speed up the development of topological thermotics. 

This Perspective is organized as follows. We first introduce the progress of topological thermal conduction in both 1D and higher dimensions. The underlying principle is the mapping between conduction equation and Schr\"odinger equation. After the introduction of convection, the thermal system is described by a non-Hermitian Hamiltonian. We then discuss the realization of thermal exceptional point and its peculiar properties. Additionally, we show the implementation of non-Hermitian topological phases in thermotics. We next discuss the topology in thermal radiation, which is classified into near-field and far-field. As for the outlook, we show some main future directions in topological thermotics, including the extension to other mass diffusion processes such as particle diffusion and plasma diffusion, and the integration of machine learning. We also discuss the potential applications of topological thermotics.

\section*{Topological thermal conduction} 

\begin{figure}[!ht]
\includegraphics[width=\linewidth]{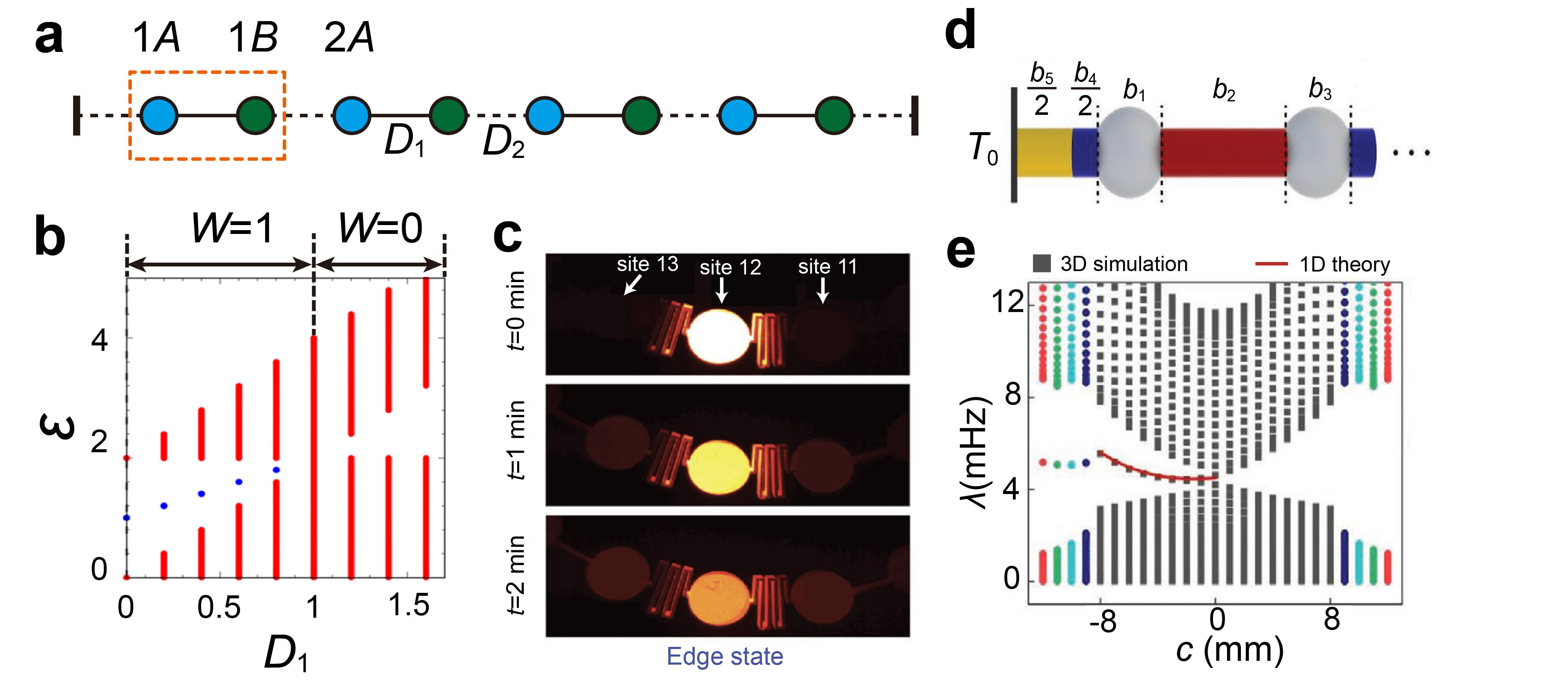}
\caption{{\bf 1D topological phase in pure conduction systems.} {\bf a}, Diffusive 1D Su--Schrieffer--Heeger model under fixed boundary conditions\textsuperscript{\cite{YoshidaSR21}}. Vertical bars at both ends denote constant-temperature heat reservoirs. Each site is labelled by ($i,\alpha$), respectively the unit cell and sublattice indices. The coupling strengths are denoted by the diffusion coefficients $D_{1}$ (solid rod) and $D_{2}$ (dashed rod). {\bf b}, Spectrum under fixed boundary conditions with $D_{2}=1$\textsuperscript{\cite{YoshidaSR21}}. $W$ is the winding number of diffusive 1D Su--Schrieffer--Heeger model. $\epsilon$ is the energy of model Hamiltonian. Blue dots represent edge states. {\bf c}, Measured temperature profiles of edge state at different moments $t$\textsuperscript{\cite{HuAM22}}. {\bf d}, Beyond the tight-binding model: the sphere--rod structure\textsuperscript{\cite{QiAM22}}. $T_{0}$ is the temperature of heat reservoir. The lengths occupied by two spheres of radius $R$ are $b_1$ and $b_3$ ($=b_1$); the lengths of the (red) intracell and (blue) intercell rods of radius $R_0$ are $b_2$ and $b_4$; The length difference between the two rods is $c=(b_{4}-b_{2})/2$. {\bf e}, The simulated eigenvalues $\lambda$ for the sphere--rod structure are compared with the theoretical results from the equivalent model (1D theory)\textsuperscript{\cite{QiAM22}}. Panels {\bf a} and {\bf b} adapted with permission from Ref.\textsuperscript{\cite{YoshidaSR21}}, Springer Nature Ltd. Panel {\bf c} adapted with permission from Ref.\textsuperscript{\cite{HuAM22}}, Wiley. Panels {\bf d} and {\bf e} adapted with permission from Ref.\textsuperscript{\cite{QiAM22}}, Wiley.}
\label{Fig2}
\end{figure}

The basic principle of topological thermal conduction is the mapping between the thermal diffusion equation and the Schr\"odinger equation. In this way, the research results of topological physics in quantum systems can be naturally transferred into classical diffusion systems. To illustrate this point, consider the diffusion equation for a continuum temperature field $T(t,x)$:
\begin{equation}
\frac{{\partial}T(t,x)}{{\partial}t}=D\frac{{\partial}^{2}T(t,x)}{{\partial}x^2},
\label{diffusion_eq}
\end{equation} 
where $D$ is the diffusion coefficient. Spatially discretizing Eq.~\ref{diffusion_eq} connects the diffusion system with the tight-binding model in condensed matter physics. The 1D Su--Schrieffer--Heeger model\textsuperscript{\cite{SuPRL79}} serves as an example (Fig.~\ref{Fig2}a)\textsuperscript{\cite{YoshidaSR21}}. The temperature field at all discretized sites can be described by the `state vector' $\mathbf{T}=(T_{1A}, T_{1B}, T_{2A}, {\cdots} T_{NB})^{\rm{T}}$, where $A, B$ denote the two sublattices, $N$ is the number of unit cells and the distance between two sites is set as the unit length. Then the heat conduction equations for all discretized sites can be simultaneously written in a matrix form
\begin{equation}
i\frac{{\partial}\mathbf{T}(t)}{{\partial}t}={H}_{\rm{1D{\;}SSH}}\mathbf{T}(t),
\label{SSH_eq}
\end{equation}
which has a similar mathematical form to the Schr\"odinger equation $i\partial_{t}{\psi}={H}{\psi}$. Here ${H}_{\rm{1D{\;}SSH}}$ can be considered as the thermal counterpart of 1D Su--Schrieffer--Heeger model:
\begin{equation}
{H}_{\rm{1D{\;}SSH}}=(-i)\left(\begin{matrix}
        D_{1}+D_{2} & -D_{1} & 0 & \cdots & 0
        \\
        -D_{1} & D_{1}+D_{2} & -D_{2} & \cdots & 0
        \\
        0 & -D_{2} & D_{1}+D_{2} & \cdots & 0
        \\
        \vdots & \vdots & \vdots & \ddots & \vdots
        \\
        0 & 0 & 0 & \cdots & D_{1}+D_{2}
        \\
        \end{matrix}\right).
\label{SSH_Hamiltonian}
\end{equation}  
This Hamiltonian differs from the original 1D Su--Schrieffer--Heeger model only by a multiplication of imaginary factor $i$ and a constant energy shift, and thus preserves the topological properties of the Su--Schrieffer--Heeger model. Therefore, a formal analogy with the Su--Schrieffer--Heeger model can be established; the main difference is that here all eigenmodes are dissipative and the pure imaginary eigenvalues determine their decay rates. As the in-gap states, the topological edge modes decay slower than the `conduction band' bulk modes, but faster than the `valence band' bulk modes (Fig.~\ref{Fig2}b). In fact, the topological edge states emerge as a robust localized mode in the temperature profile (in other words, a hot spot with suppressed heat diffusion, see Fig.~\ref{Fig2}c) in the heat diffusion process, as confirmed experimentally\textsuperscript{\cite{HuAM22}}.

Although the tight-binding model provides a clear picture and inspires the progress in the study of topological thermal conduction, it is not entirely sufficient for describing genuine heat conduction. In real heat conduction systems, long-range couplings, which are neglected in the tight-binding model, always exist. It is also more proper to describe the heat conduction as in continuous media instead of in the discretized forms. Inspired by topological photonics, the topological phenomena in heat conduction systems can be characterized by the Zak phase. For instance, a sphere--rod model with constituents of equal mass density, heat capacity, and thermal conductivity\textsuperscript{\cite{QiAM22}} (Fig.~\ref{Fig2}d) is not adequately described by the Su--Schrieffer--Heeger model, whereas a continuum model can successfully describe the system and its topological properties. When the Zak phase $\mathcal{Z}=\pi$ [$c=(b_{4}-b_{2})/2<0$)], a topological edge state emerges (Fig.~\ref{Fig2}e), confirming the bulk-edge correspondence based on the Zak phase picture.

\begin{figure}[!ht]
\includegraphics[width=\linewidth]{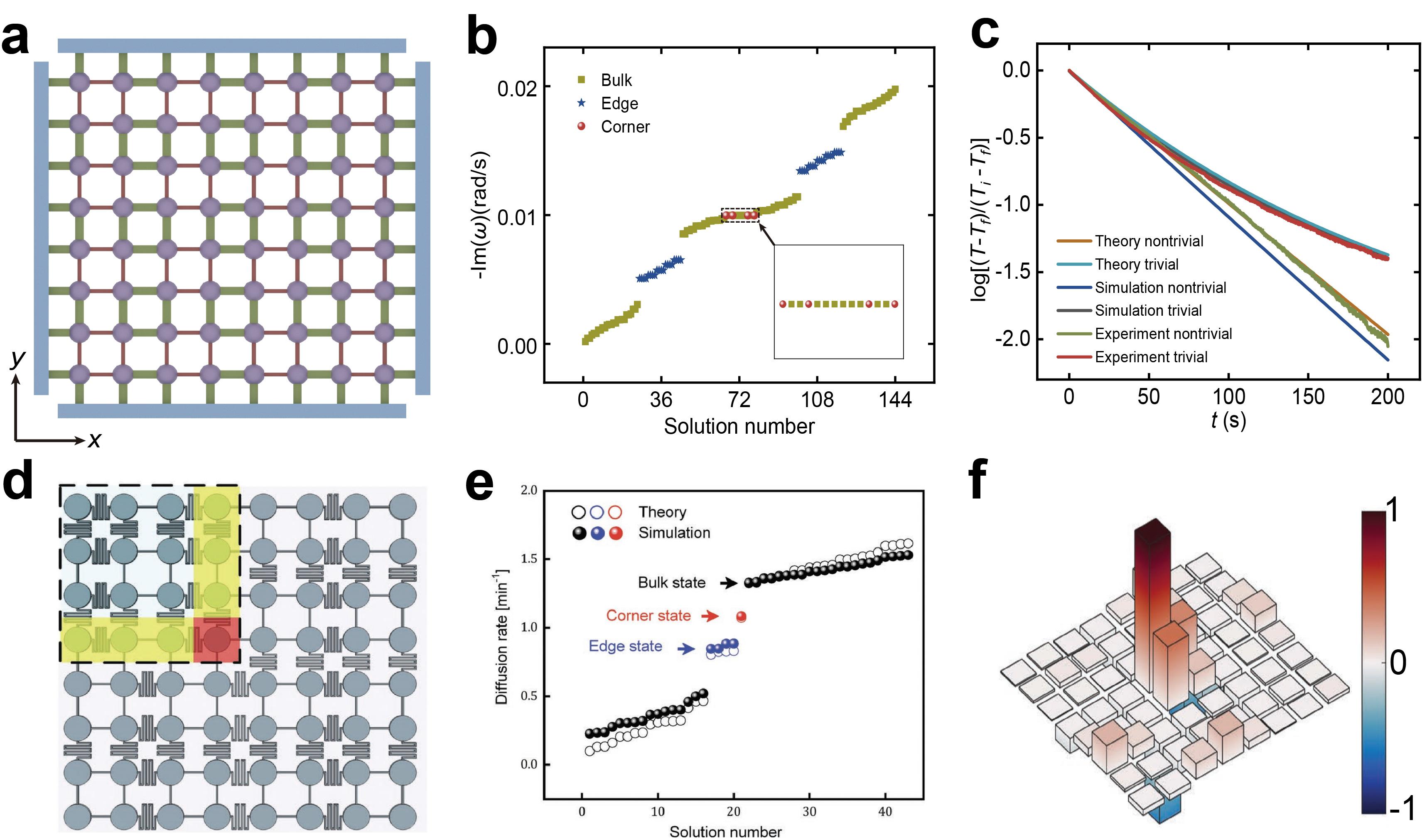}
\caption{{\bf Higher-order topological insulator in pure conduction systems.} {\bf a}, Schematic of a thermal 2D Su--Schrieffer--Heeger model without domain walls implemented as sphere--rod structure\textsuperscript{\cite{LiuPRL24}}. The boundary rods are connected with the constant temperature heat reservoirs (blue bars), that is, fixed boundary conditions. The thermal diffusivities of red thin rods are $D_1$, and green thick rods are $D_2$. {\bf b}, The fixed boundary condition spectrum of the model in part {\bf a}\textsuperscript{\cite{LiuPRL24}}. $\omega$ is the eigenvalue of model Hamiltonian. The inset is the enlarged view of four in-bulk corner states and their adjacent bulk states. {\bf c}, The theoretical, simulated and experimental normalized temperature evolution over time $t$ of corner sphere in the topological nontrivial and trivial phases\textsuperscript{\cite{LiuPRL24}}. Here, $T_i$ is the excited temperature of corner sphere and $T_f$ is the room temperature. {\bf d}, Schematic of a thermal 2D Su--Schrieffer--Heeger model with two domain walls implemented as disk--channel structure\textsuperscript{\cite{WuAM23}}. The topological nontrivial lattice, delimited by dashed lines, is located within a trivial lattice. The edge and corner states are localized in the yellow and red regions. {\bf e}, The theoretical and simulated diffusion rate spectra of the disk--channel structure\textsuperscript{\cite{WuAM23}}. {\bf f}, The experimental mode profile for the thermal corner state of the disk--channel structure\textsuperscript{\cite{WuAM23}}. Panels {\bf a}, {\bf b}, and {\bf c} adapted with permission from Ref.\textsuperscript{\cite{LiuPRL24}}, APS. Panels {\bf d}, {\bf e}, and {\bf f} adapted with permission from Ref.\textsuperscript{\cite{WuAM23}}, Wiley.}
\label{Fig3}
\end{figure}

Topological thermal conduction can be extended to higher-dimensional systems to realize systems such as diffusive higher-order topological insulators. Higher-order topological insulators have attracted much attention in condensed matter physics as well as in photonics and acoustics, primarily due to their unique properties of supporting multi-dimensional topological boundary states beyond the conventional bulk--edge correspondence\textsuperscript{\cite{BenalcazarSci17, SchindlerSA18, XieNRP21}}. For instance, a 2D higher-order topological insulator supports both 1D gapped edge states and 0D gapless corner states. In the simplest setup, such higher-order topological phases can be realized in 2D Su--Schrieffer--Heeger models\textsuperscript{\cite{BenalcazarPRB19}}. Although the 2D Su--Schrieffer--Heeger model has been studied and observed in classical wave systems\textsuperscript{\cite{XiePRL19}}, its realization and physical consequences in thermotics were reported only more recently\textsuperscript{\cite{LiuPRL24, WuAM23}}. Ref.\textsuperscript{\cite{LiuPRL24}} describes a realization of the thermal 2D Su--Schrieffer--Heeger model through a sphere--rod structure. To keep the same onsite diffusivities at all spheres, fixed boundary conditions are imposed by connecting the boundary rods with constant temperature heat reservoirs (Fig.~\ref{Fig3}a). The effective diffusive Bloch Hamiltonian for the sphere--rod structure can then be written as
\begin{equation}
{{H}}_{\rm{2D{\;}SSH}}(k_x,\,k_y)=\
(-i)\left(\begin{matrix}
        2\left(D_{1}+D_{2}\right) & -D_{1}-D_{2}e^{ik_{x}} & 0 & -D_{1}-D_{2}e^{ik_{y}}
        \\
        -D_{1}-D_{2}e^{-ik_{x}} & 2\left(D_{1}+D_{2}\right) & -D_{1}-D_{2}e^{ik_{y}} & 0
        \\
        0 & -D_{1}-D_{2}e^{-ik_{y}} & 2\left(D_{1}+D_{2}\right) & -D_{1}-D_{2}e^{-ik_{x}}
        \\
        -D_{1}-D_{2}e^{-ik_{y}} & 0 & -D_{1}-D_{2}e^{ik_{x}} & 2\left(D_{1}+D_{2}\right)\\
\end{matrix}\right),
\end{equation}
where $k_{x}$ and $k_{y}$ are the Bloch vectors along the $x$ and $y$ directions. This effective diffusive Hamiltonian has the same topological properties as the original 2D Su--Schrieffer--Heeger model. Thanks to the protection of the lattice symmetry, four thermal corner states are embedded in the bulk, as shown in the fixed boundary condition spectrum (Fig.~\ref{Fig3}b). After exciting the corner sphere, its temperature evolution follows an exponential decay in the topological nontrivial phase (Fig.~\ref{Fig3}c), the rate of which can be extracted from the fixed boundary condition spectrum. In contrast, in the trivial phase, the temperature evolution of the corner sphere deviates significantly from the exponential decay (Fig.~\ref{Fig3}c). 

Another thermal version of the 2D Su--Schrieffer--Heeger model has been demonstrated experimentally using a 2D array of aluminium disks and channels\textsuperscript{\cite{WuAM23}}; this system is inspired by prior works on topological thermotics\textsuperscript{\cite{HuAM22}}. In the setup, the varied straight and meandering channels correspond to different thermal diffusivities (Fig.~\ref{Fig3}d). The topological thermal lattice consists of $8{\times}8$ sites, with the topological nontrivial region surrounded by a trivial one. The two domain walls thus localize edge and corner states inside the lattice (Fig.~\ref{Fig3}d). Here the thermal corner state lies in the bandgap rather than the bulk, owing to the domain wall structure (Fig.~\ref{Fig3}e). The observed eigenmode profile for the corner state is pinpointed at the convergence of the two domain walls (Fig.~\ref{Fig3}f). Upon thermally exciting the domain walls according to the displayed mode profile of the corner state, the ensuing thermal relaxation of the system is governed by the corner state itself because it is an eigenmode of the system. These results have been extended to the 3D Su--Schrieffer--Heeger model\textsuperscript{\cite{ChenPRB24}}, which simultaneously supports topological surface, hinge and corner states. Furthermore, a thermal kagome lattice can also serve as an excellent example of higher-order topological insulator for thermal transport\textsuperscript{\cite{FukuiPRE23, Qiarxiv23}}.

In addition to the above, other higher-dimensional topological phases have been implemented in pure diffusion systems. An example is the helical edge modes in an honeycomb lattice inspired by the quantum spin Hall states\textsuperscript{\cite{FunayamaCP23}}. The temperature field can be given a selectable diffusion direction along the edge by suitably exciting one of the two helical edge branches. In that system, a heat pseudospin is constructed to mimic the quantum spin Hall states. Because in a honeycomb lattice such a construction is based on the six-fold ($C_6$) rotation symmetry, the robustness of the thermal edge state can be destroyed when the $C_6$ symmetry is broken\textsuperscript{\cite{FunayamaAPL23}}. Such symmetry breaking opens a gap in the edge spectrum, which can degrade the helical edge states.

The works discussed in this section are at the macroscopic scale. There are also some efforts to investigate the thermal topology in the microscale and nanoscale. At these smaller scales, the validity of the Fourier law may be compromised. Phonons play an important role in the microscale and nanoscale heat transfer, leading to numerous studies on their interplay with topology. For example, phonon thermal Hall effect, the emergence of transverse heat flow induced by a magnetic field\textsuperscript{\cite{StrohmPRL05}}, has a topological origin of the phonon bands in dielectric materials\textsuperscript{\cite{ZhangPRL10}}. How to bridge the thermal topology at the macroscale and micro- or nanoscale is an open question to be tackled in the near future. 

\section*{Topological thermal convection} 

Heat convection is another fundamental heat transfer mechanism. It is often coupled with heat conduction and offers versatile control of heat. The introduction of convection (a wave field) into heat transport not only transcends the limitations of pure conduction (a diffusion field), but also lays the foundations for new phenomena and approaches in the study of topological thermotics. Indeed, combining convection and conduction requires transforming a diffusive system from being primarily anti-Hermitian (having a purely imaginary spectrum) to generally non-Hermitian (having a complex spectrum)\textsuperscript{\cite{Liuarxiv23-1}}. Therefore, with such considerations one enters the realm of non-Hermitian physics of heat transport. Non-Hermitian Hamiltonians generally describe the dynamics of open systems coupled to a surrounding environment\textsuperscript{\cite{AshidaAP20, BergholtzRMP21}}. However, under certain conditions, such as in the presence of parity--time (PT) symmetry\textsuperscript{\cite{BenderPRL98}}, the spectrum of the system can be entirely real. Such a pseudo-Hermitian property can induce interesting features, such as unidirectional transport and enhanced sensing at the exceptional point\textsuperscript{\cite{Kato66}}. In wave systems, gain and loss correspond to the anti-Hermitian component of the effective Hamiltonian, which helps to establish the PT symmetry (Fig.~\ref{Fig4}a). Conversely, in heat transport, wave-like heat convection is modelled by Hermitian terms, whereas pure heat conduction terms remain anti-Hermitian. Balancing the heat convection in forward and backward moving fields gives rise to the emergence of anti-PT symmetry (Fig.~\ref{Fig4}a), whereby the anti-PT symmetric Hamiltonian anti-commutes with the PT operator\textsuperscript{\cite{LiSci19}}.

The archetypal design to demonstrate the anti-PT symmetry and detect exceptional points in thermal devices is the double ring model\textsuperscript{\cite{LiSci19}}. This model consists of two rings rotating in opposite directions with equal angular speed (Fig.~\ref{Fig4}b). The governing heat transfer equations of two rings are
\begin{align}
\frac{{\partial}T_{1}}{{\partial}t}=D\frac{{\partial}^{2}T_{1}}{{\partial}x^2}-v\frac{{\partial}T_{1}}{{\partial}x}+h(T_2-T_1), \\
\frac{{\partial}T_{2}}{{\partial}t}=D\frac{{\partial}^{2}T_{2}}{{\partial}x^2}+v\frac{{\partial}T_{2}}{{\partial}x}+h(T_1-T_2), 
\label{double_ring_eq}
\end{align}
where $T_{1}\,(T_{2})$ are the temperature field of upper (lower) ring, $D$ is the diffusivity of the rings,  $v$ is the rotational velocity of the rings, and $h$ is the heat exchange rate of the interlayer. After substituting a plane wave solution $\mathbf{T}=(T_{1}, T_{2})^{\rm{T}}=(A_{1}, A_{2})^{\rm{T}}{\cdot}e^{i(kx-{\omega}t)}$ (where $A_{1}\,(A_{2})$ are the amplitudes, $k$ is the wave number, and $\omega$ is the eigenfrequency) into the above governing equations, these two equations can be written in the same form as the Schr\"odinger equation
\begin{equation}
i\frac{{\partial}\mathbf{T}}{{\partial}t}={H}\mathbf{T},
\label{double_ring_eq_2}
\end{equation}
where the effective anti-PT symmetric Hamiltonian of double ring model is
\begin{equation}
{{H}}=\left(\begin{matrix}
            -i(k^{2}D+h)+kv & ih
            \\
            ih & -i(k^{2}D+h)-kv
            \end{matrix}\right).
\label{double_ring_Ham}
\end{equation}  
The corresponding eigenvalues are 
\begin{equation}
\omega_{\pm}=-i\left[(k^{2}D+h){\pm}\sqrt{h^{2}-k^{2}v^{2}}\right]
\label{double_ring_eigenvalue}
\end{equation}  
with an exceptional point at $h^{2}=k^{2}v_{\rm{EP}}^2$. The two decay rate branches converge at the exceptional point and nonzero eigenfrequencies emerge with increasing $v$ . The temperature profile in the rings stabilizes after a certain time in the anti-PT unbroken phase (left panel of Fig.~\ref{Fig4}c), whereas it drifts steadily due to the nonzero eigenfrequency in the anti-PT broken phase (right panel of Fig.~\ref{Fig4}c). The key mechanism behind this phenomenon is the competition between coupled thermal convection and conduction. The real coupling between counter-rotating rings has also been realized experimentally, thus demonstrating the PT symmetry in diffusive systems\textsuperscript{\cite{CaoSA24}}. By increasing the number of rings from two to four, one can achieve a thermal third-order exceptional point with augmented robustness against perturbations\textsuperscript{\cite{CaoESEE20}}.  

The existence of exceptional points and their topological properties are a topic of current investigation in non-Hermitian physics\textsuperscript{\cite{DopplerNat16, XuNat16, YoonNat18}}. Moving adiabatically along a closed path encircling an exceptional point results in the transition from the initial eigenstate to a different one. A constrained parameter space in thermotics can be addressed by considering a multi-ring system\textsuperscript{\cite{XuPRL21}} (Fig.~\ref{Fig4}d). This structure integrates two orthogonal sets of counter-rotating rings corresponding to a broad parameter space including two spectral exceptional points. When the evolution path in the convection parameter space encircles two exceptional points, the temperature field returns to its initial distribution, which signals a non-Hermitian thermal topology. Here, a closed adiabatic path traverses a complete Riemann surface to return to its initial state. The geometric phase generated by encircling the single thermal exceptional point of the double ring model has also been analyzed\textsuperscript{\cite{XuIJHMT21}}. When the time-varying rotating velocity encircles the thermal exceptional point once, a $\pi$ geometric phase is accumulated in the temperature field (Fig.~\ref{Fig4}e). The chiral behaviour around an exceptional point in optics has inspired innovations like asymmetric wave propagation and one-way invisibility in tailored whispering-gallery-mode resonators\textsuperscript{\cite{WangNP20, PengPNAS16}}. Following this approach, the chiral properties of thermal exceptional points in the presence of thermal perturbations have been investigated\textsuperscript{\cite{XuPRL23}}. In the anti-PT broken phase, the temperature distribution around a thermal exceptional point turns asymmetric regardless of the direction or magnitude of advection (Fig.~\ref{Fig4}f). In contrast, in the anti-PT unbroken phase or precisely at the exceptional point, the temperature distribution hints at a non-chiral heat flow. Therefore, chirality offers a means to manipulate heat flows at will, which causes an anisotropic thermal conductivity.

Non-Hermitian physics also enriches topological physics, and thermal metamaterials provide an excellent platform to observe non-Hermitian topological effects.Modulating convection makes it possible to realize a 1D non-Hermitian topological insulator in thermotics\textsuperscript{\cite{XuNP22}}. The starting structure is a planar lattice with four periodic convection units. For practical implementation, the two ends of one unit are connected to form a ring (Fig.~\ref{Fig5}a). For a particular convection arrangement, the system is topologically nontrivial with a pair of edge states (Fig.~\ref{Fig5}b). The thermal profile for this arrangement is robust and stationary, which is a signature of topological edge state. Furthermore, a suitable combination of trivial and nontrivial lattices generates a topological interface state, showing remarkable capability in controlling the thermal field. Replacing real convection with imitated advection by graded conduction metadevices\textsuperscript{\cite{XuNSR23}} was proposed as a means to produce a non-reciprocity induced non-Hermitian skin effect\textsuperscript{\cite{CaoCP21}}. An experimental demonstration of this effect has been reported for the coupled ring chain structure\textsuperscript{\cite{LiuSB24}}. Finally, incorporating real convection resulted in a phase-locking diffusive skin effect\textsuperscript{\cite{CaoCPL22}}. The parameter modulation technique is flexible enough to realize non-Hermitian diffusive quasicrystals, which shows a unique phenomenon of multiple localization centres\textsuperscript{\cite{LiuPRAP24}}.

In 2D systems, a notable advance is the realization of a quadrupole topological insulator for heat transport\textsuperscript{\cite{XuNC23}}. The basic structure comprises a convective thermal lattice characterized by discrete sites. The tunable advections ${\it{\Omega}}_{{\uppercase\expandafter{\romannumeral1}}/{\uppercase\expandafter{\romannumeral2}}}$ are applied to each site, yielding a unit cell of four sites. To achieve both positive and negative nearest-neighbour couplings simultaneously, the design incorporates two types of tilted connections with opposite isotherms. Thanks to the coexisting Hermitian advection and non-Hermitian thermal coupling modulation, a robust in-gap corner state manifests itself in both the real and imaginary spectra. The corresponding temperature field shows a corner-localized mode (Fig.~\ref{Fig5}c). This finding extends the conventional quadrupole topological insulator from real spectrum to complex spectrum. A separate work proposes the implementation of a non-Hermitian Chern insulator of heat with chiral edge state in a hexagonal array\textsuperscript{\cite{XuEPL21}}. Different phases with various Chern numbers can be realized, where the edge states can be tuned by topological transitions (Fig.~\ref{Fig5}d). In addition, the higher-order non-Hermitian skin effect in heat transport has been realized in a square array through imitated advection\textsuperscript{\cite{HuangCPL23}}.

Moving to even higher dimensions, a 3D gapless Weyl exceptional ring can be realized in topological thermotics\textsuperscript{\cite{XuPNAS22}}. The Weyl exceptional ring, a curve made of infinite exceptional points, emerges from the Weyl point in Hermitian topology\textsuperscript{\cite{XuPRL17}}. It has been demonstrated in topological photonics\textsuperscript{\cite{CerjanNP19}} and topological acoustics\textsuperscript{\cite{LiuPRL22}}. The thermal Weyl exceptional ring is observed for a basic model consisting of a central strip and four counter-convective components (Fig.~\ref{Fig5}e). The virtual parameter space of this structure makes it possible to adiabatically encircle the Weyl exceptional ring. During the process, the temperature profile remains stationary while the phase changes by $\pi$, as expected due to the nontrivial topology (Fig.~\ref{Fig5}f). 

\section*{Topological thermal radiation} 

As the third mechanism of heat transfer, heat radiation is the process by which objects emit energy as electromagnetic waves, a phenomenon largely controlled by their temperature. This section introduces the topology in thermal radiation. When focusing on thermal radiation we are led to distinguish between near-field and far-field radiation. Near-field radiation becomes prominent when the source--observer distance closely matches or is even shorter than the emitted wavelength\textsuperscript{\cite{BiehsRMP21, TianCPL23, GeCPL23}}. It has attracted widespread interest for its ability to surpass the blackbody limit and works at the micro or nano scale. The fundamental theoretical framework for near-field radiation is fluctuation electrodynamics. 

Near-field radiation among nanoparticles has been related to the emergence of certain topological effects. In a 1D Su--Schrieffer--Heeger chain of indium antimonide nanoparticles, in the topologically nontrivial phase of radiative heat flux, edge modes, despite their strong localization, govern radiative heat transport thanks to the long-range coupling between the first and last chain particles\textsuperscript{\cite{OttPRB20}}. In a 2D honeycomb lattice, near-field radiative heat transport predominantly relies on the edge rather than on the bulk modes\textsuperscript{\cite{OttIJHMT22}}. To better understand this mechanism, a general expression for the near-field energy density within any environment\textsuperscript{\cite{OttPRB21}} can be used. In the example of a 2D Su--Schrieffer--Heeger lattice, the spectral energy density with corner mode eigenfrequency is notably amplified in the nontrivial phase. In addition, near-field radiative heat transfer is influenced by residual surface charges\textsuperscript{\cite{LuoMTP23}}, an analogue of 3D topological insulators with surface states. To experimentally assess topological edge modes, researchers have combined near-field thermal imaging with far-field radiation, so as to capture the direct thermal emission spectrum of a plasmonic Su--Schrieffer--Heeger chain of indium antimonide nanoparticles\textsuperscript{\cite{HerzAPL22}}. However, these studies have predominantly focused on the static regime of the near-field radiative thermal topology, omitting to address transient or steady-state heat transfer. To fill this gap, the energy balance equation has been linearized to extract the temperature time dependence along the chain\textsuperscript{\cite{NikbakhtPRB23}}. The ensuing response matrix reflects the topological nature of the system. These studies suggest new promising strategies to better manipulate near-field radiation.

Conversely, far-field radiation is typically observed when the wavelength of the emitted radiation is significantly smaller than the source--observer distance, a situation well described by the Stefan--Boltzmann law. According to the Rosseland diffusion approximation\textsuperscript{\cite{Rosseland31, XuESEE20, XuPRAP20}}, far-field radiative heat flux is proportional to the temperature gradient and the proportionality coefficient is related to the cube of temperature. So we can envisage that far-field radiation could be treated by means of nonlinear temperature dependent terms being introduced into the anti-Hermitian pure conduction and non-Hermitian conduction--convection.

\section*{Outlook} 

Inferring from the development of topological photonics and acoustics, we anticipate a vigorous growth of topological thermotics in the near future. There are several directions that are promising and anticipated.

The thermal diffusion equation has a similar form with governing equation of other diffusion processes. Therefore, the studies of topological thermotics can be transplanted into other diffusion processes to investigate their topology. Next we will show two examples. 

The first example is topological particle dynamics. Particle dynamics is a common mass transport process driven by concentration gradients\textsuperscript{\cite{HanggiRMP09}}. The central equation of particle diffusion is Fick's law, which has a same form with the thermal diffusion equation. Based on this similarity, concepts from topological thermotics, such as the geometric phase in topological particle diffusion (Box 1), may inspire research in  topological particle dynamics\textsuperscript{\cite{XuPRE21}}. 

The second example is topological plasma physics. Plasma\textsuperscript{\cite{Lieberman05, Chen16}}, the ``fourth state of matter", exhibits collective transport properties characterized by nonlinear diffusion\textsuperscript{\cite{HuangIEEE15}}. Recent years have witnessed the application of transformation theory to plasma transport equations, albeit under stringent approximations\textsuperscript{\cite{ZhangCPL22}}. This simplified plasma transport equation has the same form as the thermal conduction--convection equation. Thus exotic topological phases in plasma physics can be realized through borrowing the wisdom of topological thermotics (Box 2), such as the plasma version of phase-locking skin effect and non-Hermitian quasicrystal\textsuperscript{\cite{LiuCPL23}}. Although topology has been applied in plasma physics to interpret certain distinctive phenomena\textsuperscript{\cite{GaoNC16, ParkerPRR20, ParkerPRL20, FuNC21}}, these studies predominantly view plasma as a fluid driven by electromagnetic forces. Topological plasma diffusion can effectively complement the topological plasma physics based on magnetohydrodynamics. However, the theory of diffusive topological plasma transport warrants further investigation due to it having too many simplifications. In addition, the phenomenon of anti-diffusion in topological plasma transport is readily accessible within the diffusive paradigm\textsuperscript{\cite{YokoiarXiv23}}.

A prevailing direction in topological thermotics is the realization of additional thermal topological states. The mapping between the pure conduction equation and the Schr\"odinger equation renders topological models from condensed matter physics accessible to thermal diffusion. Various topological models can be designed to realized in thermal lattices, such as kagome\textsuperscript{\cite{YinNat20, ZhengNat22, ZhongNat23}}, quasicrystal\textsuperscript{\cite{HuangPRL18}}, moir\'e lattice\textsuperscript{\cite{WangNat20, LiuNat21-2, WangNat22, UriNat23, ZengNat23, LiNC24}}, and fractal structures\textsuperscript{\cite{BiesenthalSci22}}. The Majorana zero mode could also be implemented in artificial thermal metamaterials, thus avoiding the controversies persisting in condensed matter physics\textsuperscript{\cite{YazdaniSci23, ValentiniSci21, LiNat22}}. In addition, the nonlinearity in topological thermotics can be effectively controlled via temperature-dependent thermal conductivity\textsuperscript{\cite{LiPRL15, ShenPRL16}}. Going beyond the tight-binding model is also possible in topological thermotics\textsuperscript{\cite{QiAM22}}. With the help of convection, one could dig deeper into the non-Hermitian thermal topology. Indeed, by enlarging the virtual convection space, non-Hermitian topological phases in higher dimensions, such as three-dimensional exceptional lines\textsuperscript{\cite{YangPRB19}}, might become accessible. Furthermore, the easy tunability of heat convection could bring the braiding of non-Hermitian topological bands to reality\textsuperscript{\cite{WangNat21}}. Beyond non-Hermitian physics, the implementation of Floquet dynamics through spatiotemporal parameter modulation is another promising direction of topological thermotics\textsuperscript{\cite{Yinelight22}}.   

The introduction of machine learning is a promising direction in the field of topological thermotics. On one hand, machine learning method has been adopted to classify different topological phases and calculate their topological invariants\textsuperscript{\cite{ZhangPRL17, ZhangPRL18, LongPRL20, ScheurerPRL20}}. It has also been proposed that machine learning can aid ab initio calculations in predicting real topological materials\textsuperscript{\cite{ClaussenPRB20, CaoPRM20, LiuPRM21, SchlederAPR21, AndrejevicAM22, MaNL23}}. On the other hand, machine learning techniques have provided an alternative approach to design a variety of thermal metamaterials efficiently\textsuperscript{\cite{ZhuCR24, HuPRX20, HuNE20, JiIJHMT22, JinAM24}}. Therefore, we believe that the integration with machine learning could assist research in topological thermotics. 

Finally, topological thermotics lends itself to a variety of practical applications. Manipulating thermal edge states has been predicted to aid efficient heat dissipation in cooling of engines and electronic chips. In addition, this heat management is localized without affecting the neighbourhood\textsuperscript{\cite{HuAM22, QiAM22, LiuPRL24, WuAM23}}, which means it can be used as a novel way for heat trapping of energy. The robustness of thermal edge states against defects and disorder may facilitate the storage and transmission of thermal information. The manipulation of thermal interface states at will could inspire new heat control technology\textsuperscript{\cite{XuNP22}}. Topological thermal sensing can be achieved by exploiting the high sensitivity of diffusive skin effects to boundary conditions\textsuperscript{\cite{CaoCP21, BudichPRL20}}. Topological thermotics can find potential applications even in astronautics, although in vacuum heat convection is suppressed. In this case, topological heat conduction and radiation would play a dominant role. For example, the heat inside a space suit could be collected by topological heat conduction and dissipated outside by topological heat radiation. The generation of electric power in a spacecraft could also result from the combination of topological heat conduction and radiation. Topological thermotics may also help enhance the thermoelectric\textsuperscript{\cite{MaoSci19, RenCPL23}} and radiative cooling\textsuperscript{\cite{RamanNat14, ZhaiSci17, LinSci23, ZhaoSci23}} efficiency.

In brief, topological thermotics has far-reaching implications in both fundamental research and practical applications, which will receive more attention from researchers in metamaterials, condensed matter physics, and thermal science. 

\section*{Acknowledgments}
The authors are indebted to Dr. Fubao Yang and Dr. Liujun Xu for their insightful comments and suggestions on this review, and extend their gratitude to Prof. Ruibao Tao for his invaluable and encouraging discussions regarding the study of topology in thermal metamaterials. J. H. was supported by the National Natural Science Foundation of China under Grants No. 12035004 and No. 12320101004, and the Innovation Program of the Shanghai Municipal Education Commission under Grant No. 2023ZKZD06. J.-H. J. was supported by the National Natural Science Foundation of China (Grant No. 12125504), and the ``Hundred Talents Program'' of the Chinese Academy of Sciences. F. M. was supported by the National Natural Science Foundation of China (Grant No. 12350710786).

\section*{Author contributions}
Z.L. researched data for the article. Z.L., F.M., J.-H.J. and J.H. contributed substantially to discussion of the content. Z.L., P.J., M.L. and C.W. wrote the article. J.-H.J. and J.H. reviewed and/or edited the manuscript before submission.

\section*{Competing interests}
The authors declare no competing financial interests.

\clearpage
\newpage

\section*{Figure captions}

\clearpage
\newpage

\clearpage
\newpage

\clearpage
\newpage
\begin{figure}[!ht]
\includegraphics[width=\linewidth]{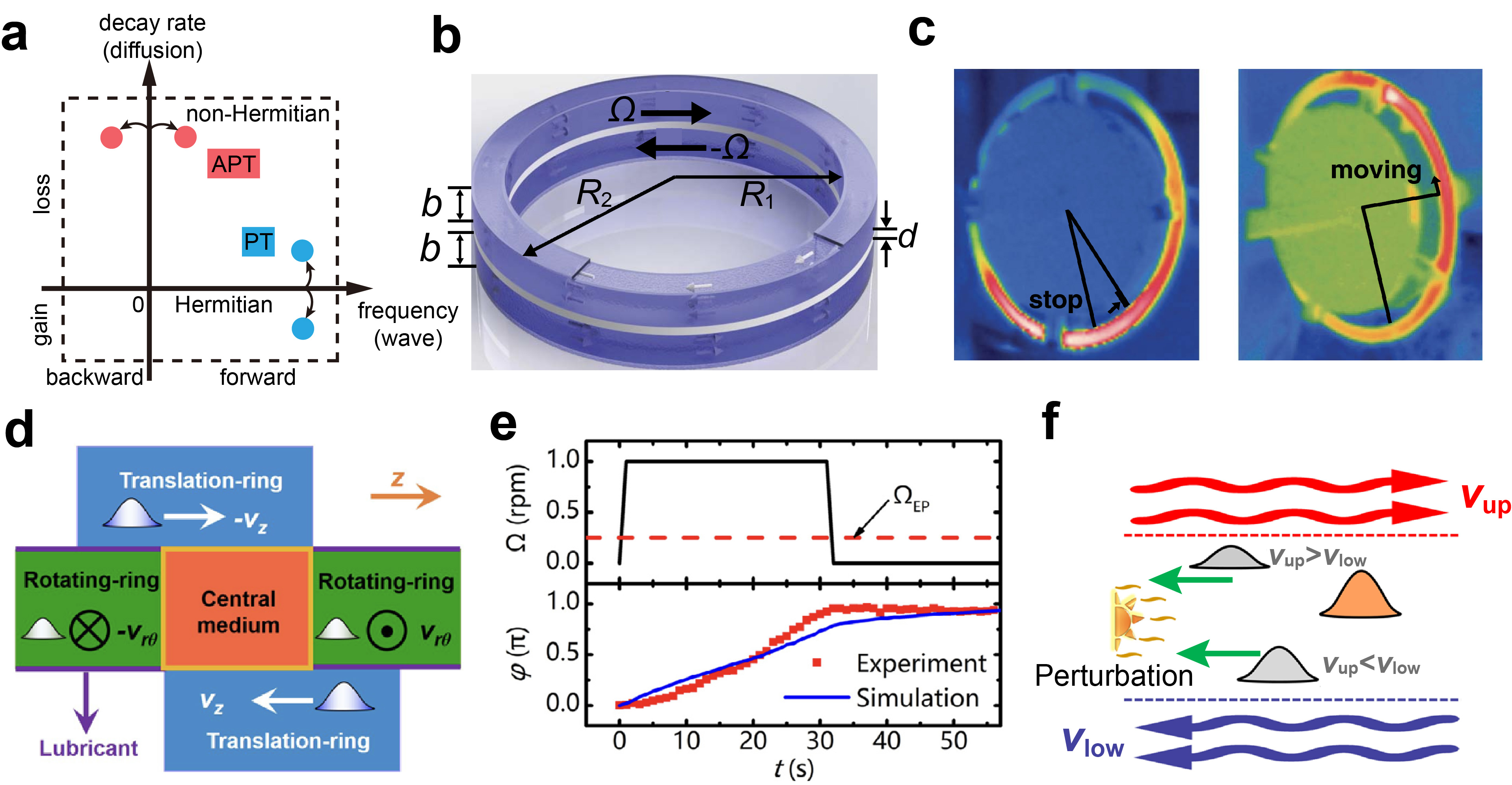}
\caption{{\bf Exceptional point and its properties in conduction-convection systems.} {\bf a}, Two strategies to demonstrate the parity--time (PT)-associated phenomena: incorporating gain and loss in wave systems (blue dots) and introducing forward and backward convection fields in diffusive systems (red dots)\textsuperscript{\cite{LiSci19}}. {\bf b}, Schematic of a double ring model\textsuperscript{\cite{LiSci19}}. {\bf c}, Experimental temperature fields for the double ring model in the anti-PT unbroken (left) and broken (right) phases\textsuperscript{\cite{LiSci19}}. {\bf d}, Cross section of the multiple ring system with two orthogonal pairs of advective modulations\textsuperscript{\cite{XuPRL21}}. $v_{r\theta}$ is the rotating velocity in the $r\theta$ plane. $v_{z}$ is the translational velocity along the $z$ direction. {\bf e}, The experimental and simulated $\pi$ geometric phase (lower panel) when the time-varying rotating velocity encircles the thermal exceptional point once (upper panel)\textsuperscript{\cite{XuIJHMT21}}. $\Omega$ is the rotating angular velocity of rings and $\Omega_{EP}$ is the one at the exceptional point. $\phi$ is the generated geometric phase. {\bf f}, Schematic of chiral heat transport under constant thermal perturbation and advections\textsuperscript{\cite{XuPRL23}}. $v_{up}$ and $v_{low}$ are the advection velocities at the upper and lower boundaries respectively. Two wave-like advection flows are imposed at the upper and lower boundaries. The eigenstates (gray wave packets) exhibit the same directions (dark grey arrows) regardless of the magnitude of upper and lower advection. Panels {\bf a}, {\bf b}, and {\bf c} adapted with permission from Ref.\textsuperscript{\cite{LiSci19}}, AAAS. Panel {\bf d} adapted with permission from Ref.\textsuperscript{\cite{XuPRL21}}, APS. Panel {\bf e} adapted with permission from Ref.\textsuperscript{\cite{XuIJHMT21}}, Elsevier. Panel {\bf f} adapted with permission from Ref.\textsuperscript{\cite{XuPRL23}}, APS.}
\label{Fig4}
\end{figure}

\clearpage
\newpage
\begin{figure}[!ht]
\includegraphics[width=\linewidth]{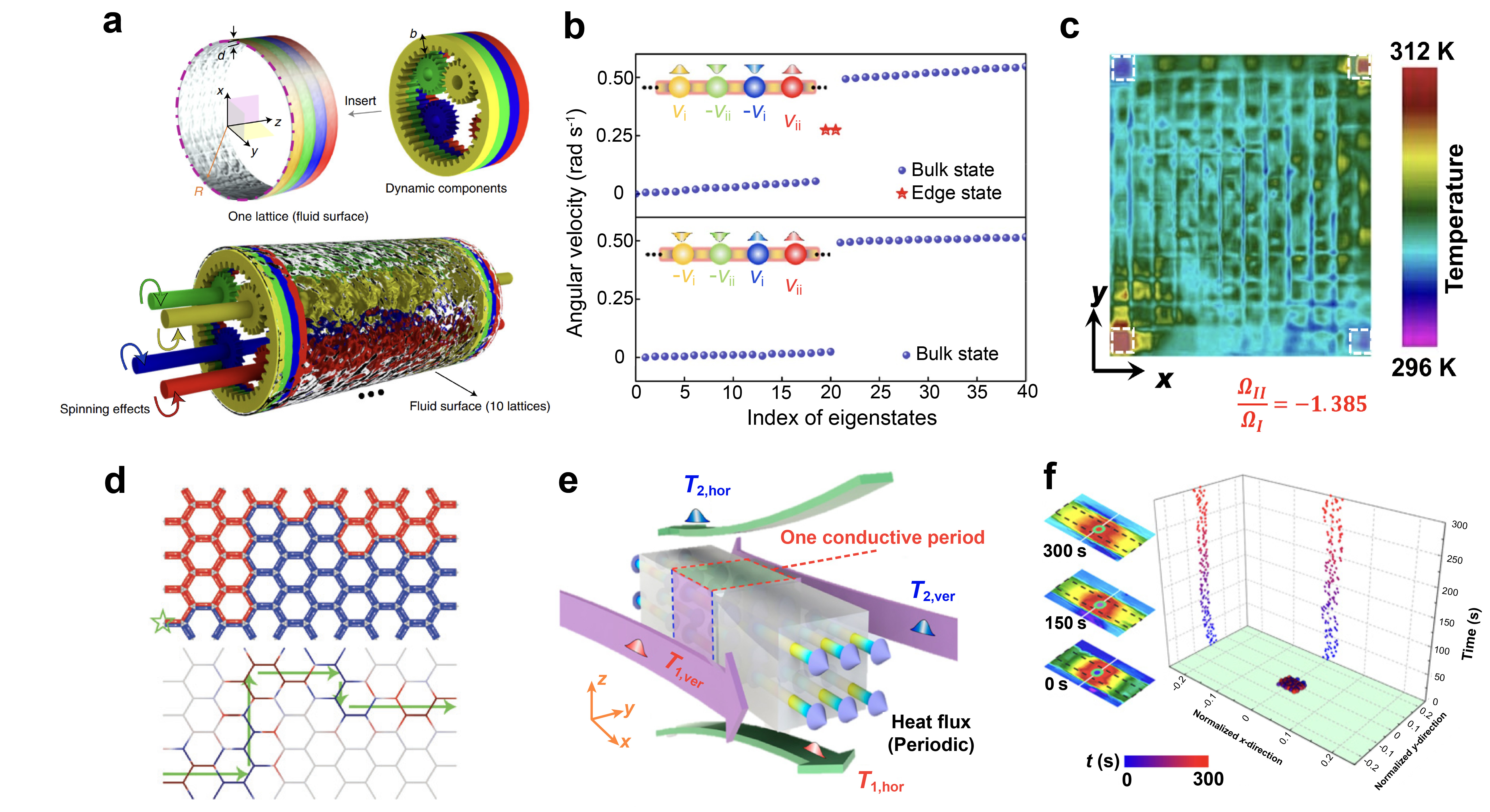}
\caption{{\bf Non-Hermitian topology in conduction-convection systems.} {\bf a}, Schematic of the coupled ring chain structure\textsuperscript{\cite{XuNP22}}. {\bf b}, The eigenvalue spectra of the nontrivial and trivial coupled ring chain structures\textsuperscript{\cite{XuNP22}}. {\bf c}, Experimental temperature field of thermal corner state induced by advection\textsuperscript{\cite{XuNC23}}. {\bf d}, Thermal interface state in the non-Hermitian Chern insulators with clockwise and counterclockwise velocities\textsuperscript{\cite{XuEPL21}}. The star denotes the position of periodic temperature
source. The arrows show the direction of temperature propagation. {\bf e}, Structure supporting a diffusive Weyl exceptional ring with four counter advection flows on orthogonal surfaces\textsuperscript{\cite{XuPNAS22}}. {\bf f}, The measured time ($t$) evolution of the maximum temperature point in the $x-y$ plane as the integration surface encircles the full Weyl exceptional ring\textsuperscript{\cite{XuPNAS22}}. Temperature profiles at different times are shown in the inset. Panel {\bf a} and {\bf b} adapted with permission from Ref.\textsuperscript{\cite{XuNP22}}, Springer Nature Ltd. Panel {\bf c} adapted from Ref.\textsuperscript{\cite{XuNC23}} under a Creative Commons licence CC BY 4.0. Panel {\bf d} adapted with permission from Ref.\textsuperscript{\cite{XuEPL21}}, IOP Publishing. Panel {\bf e} and {\bf f} adapted from Ref.\textsuperscript{\cite{XuPNAS22}} under a Creative Commons licence CC BY 4.0.}
\label{Fig5}
\end{figure}

\clearpage
\newpage
\begin{figure}[!ht]
\includegraphics[width=0.5\linewidth]{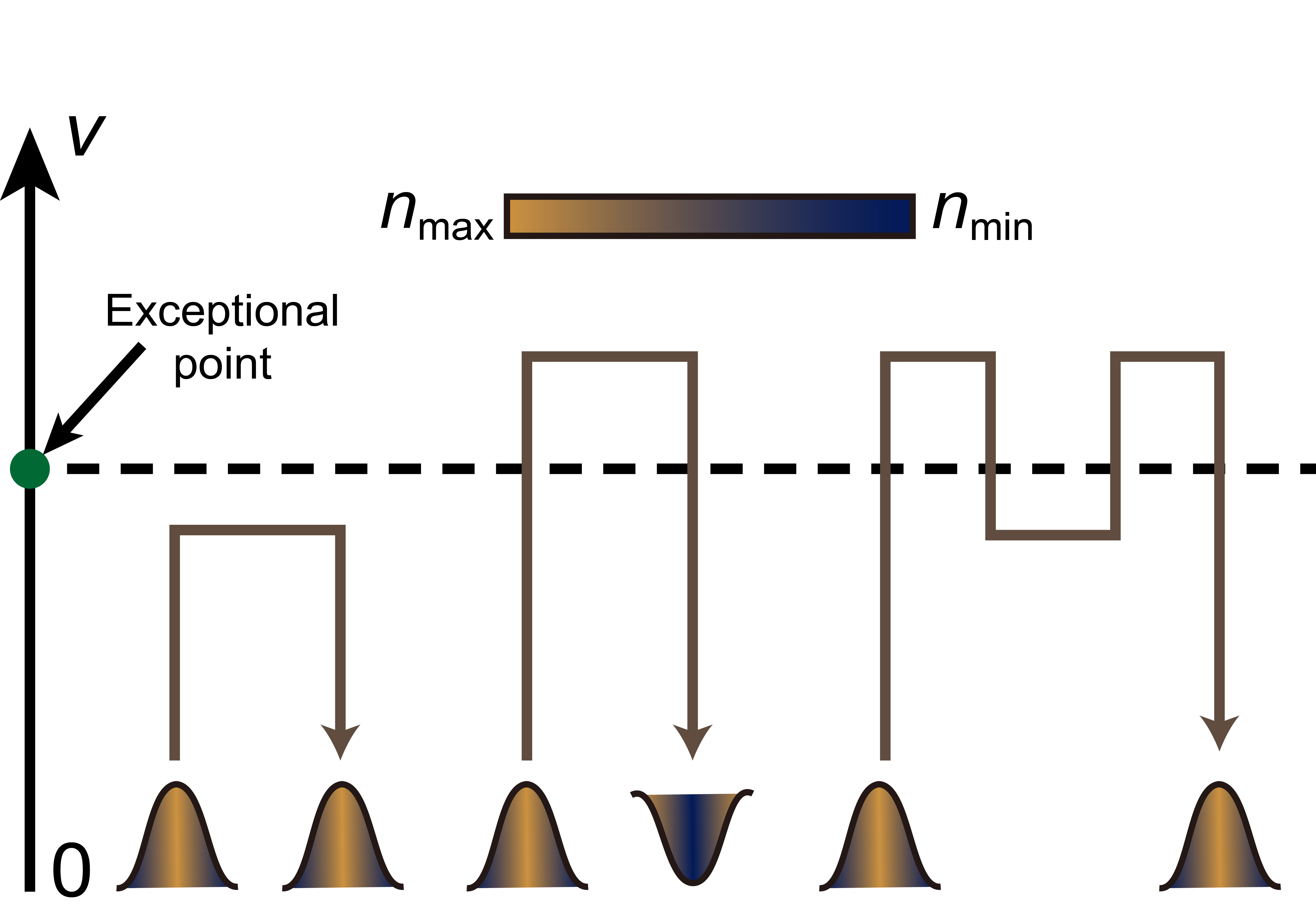}
\caption{Box 1}
\label{Box1}
\end{figure}

\clearpage
\newpage
\begin{figure}[!ht]
\includegraphics[width=0.5\linewidth]{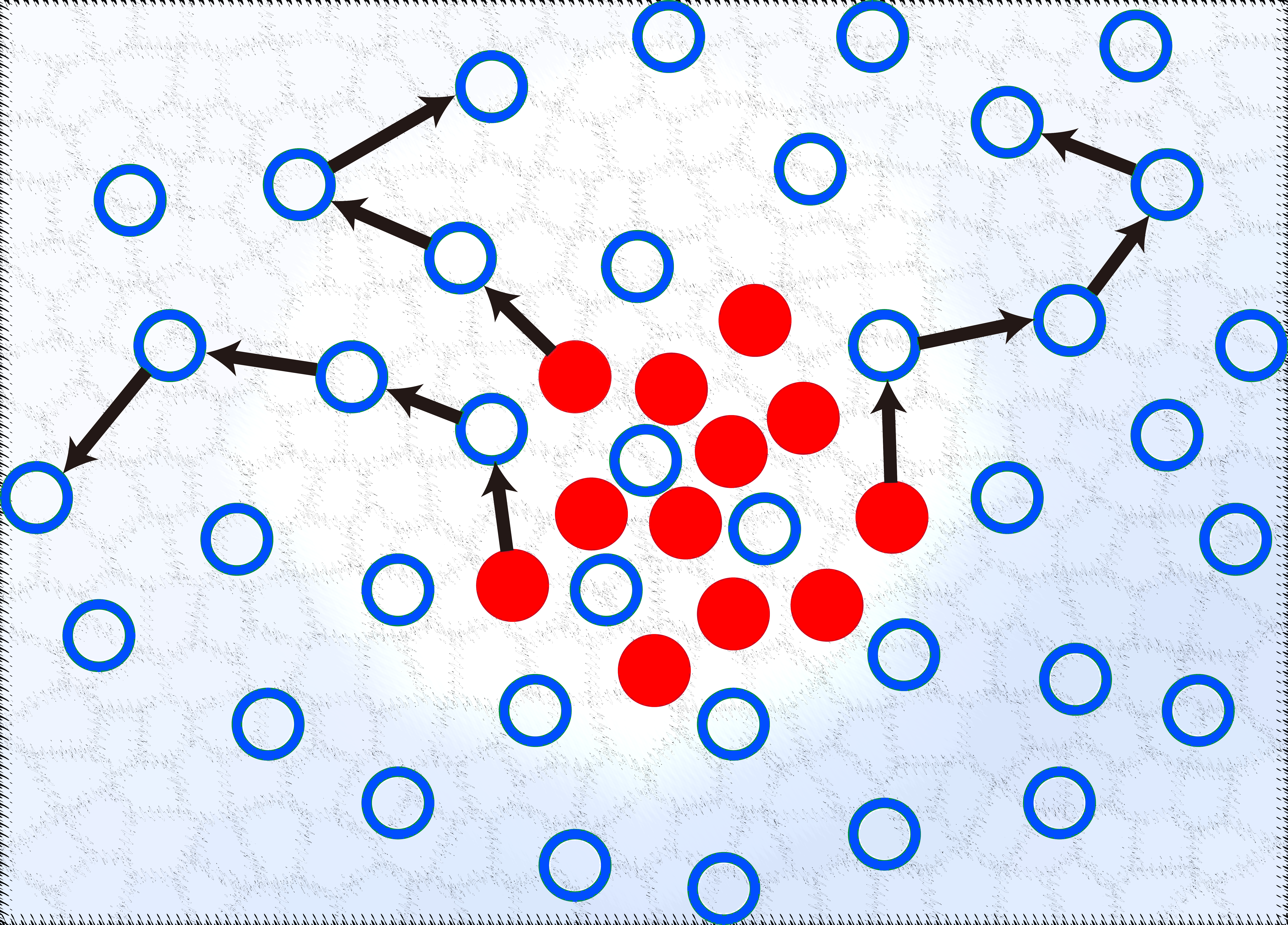}
\caption{Box 2}
\label{Box2}
\end{figure}

\section*{Box 1: Geometric phase in topological particle dynamics}

An intuitive approach to topological particle dynamics involves taking insights from topological thermotics due to the form-similarity between governing equations. As an example, the geometric phase has been realized in particle diffusion\textsuperscript{\cite{XuPRE21}}, which is inspired by the thermal exceptional point\textsuperscript{\cite{LiSci19, XuIJHMT21}}. The basic structure is the double ring model consisting of two equal but counter-rotating rings filled with particles, separated by a stationary interlayer (a similar structure in Fig.~\ref{Fig4}b). Like with thermotics, there is an exceptional point in the anti-parity--time (anti-PT) symmetric Hamiltonian of double ring model. Evolving along a closed trajectory in the parameter space, the density field accumulates an extra geometric phase due to the topological nature of exceptional point\textsuperscript{\cite{DopplerNat16, XuNat16, YoonNat18}}. In the double ring model, the ring rotation speed $v$ is continuously varied to make the state evolve adiabatically. When the closed trajectory encircles the exceptional point only once, an extra $\pi$ geometric phase can be generated in the density field distribution $n$ (as shown in Fig.~\ref{Box1}). However, when the closed trajectory excludes the exceptional point or encircles it twice, the density profile reverts to its initial state, exhibiting respectively a 0 or $2\pi$ geometric phase (as shown in the figure). 

\section*{Box 2: Topological plasma diffusion}

Diffusion processes play an important role in the plasma transport because of the density gradient in the real plasma\textsuperscript{\cite{Lieberman05, Chen16}}. For example, in the weakly ionized plasma, diffusion behaviour attributes to a nonuniform distribution of ions and electrons (red solid dots in Fig.~\ref{Box2}) which collide frequently with the dense background neutral atoms (blue hollow dots in Fig.~\ref{Box2}). By disregarding gaseous reactions and advection terms, the transport equation for charged particles in plasma can be reduced to a diffusion--migration equation, which has a similar form to the conduction--convection equation\textsuperscript{\cite{ZhangCPL22}}. Therefore, investigations of topological plasma diffusion can refer to topological thermotics. 

Diffusive non-Hermitian topological phases in plasma transport have been predicted\textsuperscript{\cite{LiuCPL23}}. The coupled ring chain structure was adapted to emulate the tight-binding model in condensed matter physics (a similar structure in Fig.~\ref{Fig5}a). Based on this structure, two topological phases were realized: One is the phase-locking skin effect and the other is the non-Hermitian quasicrystal with a complex quasiperiodic onsite potential. This diffusive paradigm may motivate the realization of more plasma topological states.

\end{document}